\documentclass[twocolumn,showpacs,amsmath,prb]{revtex4}

\usepackage{graphicx}
\usepackage{color}

\newcommand{\exciting}{{\usefont{T1}{lmtt}{b}{n}exciting}}

\begin{document}

\title{Accurate all-electron $G_0W_0$ quasiparticle energies employing the full-potential augmented planewave method}

\author{Dmitrii Nabok}
\email{dmitrii.nabok@physik.hu-berlin.de}
\author{Andris Gulans}
\email{andris.gulans@physik.hu-berlin.de}
\author{Claudia Draxl}
\email{claudia.draxl@physik.hu-berlin.de}
\homepage[]{http://sol.physik.hu-berlin.de}
\affiliation{Physics Department and IRIS Adlershof, Humboldt-Universit\"at zu Berlin, Zum Gro\ss en Windkanal 6, D-12489 Berlin}
\affiliation{European Theoretical Spectroscopy Facility (ETSF)}

\date{\today}

\begin{abstract}
The $GW$ approach of many-body perturbation theory (MBPT) has become a common tool for calculating the electronic structure of materials. However, with increasing number of published results, discrepancies between the values obtained by different methods and codes become more and more apparent. For a test set of small- and wide-gap semiconductors, we demonstrate how to reach the numerically \emph{best} electronic structure within the framework of the full-potential linearized augmented planewave (FLAPW) method. We first evaluate the impact of local orbitals in the Kohn-Sham eigenvalue spectrum of the underlying starting point. The role of the basis-set quality is then further analyzed when calculating the $G_0W_0$ quasiparticle energies. Our results, computed with the \exciting{} code, are compared to those obtained using the projector-augmented planewave (PAW) formalism, finding overall, good agreement between both methods. We also provide data produced with a typical FLAPW basis set as a benchmark for other $G_0W_0$ implementations.
\end{abstract}

\pacs{}

\maketitle

\section{\label{sec:intro}Introduction}

The $GW$ approach within many-body perturbation theory (MBPT) \cite{Hedin1965} has become a standard tool for studying single-particle excitations in condensed-matter physics.\cite{Aryasetiawan1998,Aulbur1999,Onida2002} Owing to increasing computing power and advances in algorithms, this method has been applied for a wide range of problems that includes band structures of molecules,~\cite{VanSetten2015} two-\cite{Rasmussen2015} and three-dimensional~\cite{Schilfgaarde2006,Huser2013} crystalline materials, and even molecules in aqueous solutions~\cite{Opalka2015} and water/semiconductor interfaces.~\cite{Govoni2015}

Despite the overall success of the $GW$ approach, there are issues of both theoretical and numerical nature.
These issues begin with the question how to carry out $GW$ calculations in practice. The original prescription of Hedin\cite{Hedin1965} implies a self-consistent calculation of quasiparticle (QP) energies, while the modern pragmatic approach known as single-shot $GW$, or simply $G_0W_0$, abandons this self-consistency and is applied as a perturbative correction to Kohn-Sham (KS)~\cite{Hybertsen1986} or Hartree-Fock (HF) results. The $G_0W_0$ approximation is particularly attractive, because it requires a reduced computational effort compared to the self-consistent $GW$, yet $G_0W_0$ quasiparticle gaps are in a good agreement with experiment.~\cite{Schilfgaarde2006,Klimes2014a} Unfortunately, $G_0W_0$ has a dependence on the KS or HF reference~\cite{Rostgaard2010,Bruneval2013,Rinke2008} what is known as the {\it starting-point problem}. This ambiguity can be resolved by using the reference that minimizes the $G_0W_0$ correction.~\cite{Schilfgaarde2006} The $GW$ approach as such neglects all vertex contributions to the self-energy, and there have been attempts to restore them.~\cite{Gruneis2014,Chen2015} Each conceptual advance is aimed at approaching the exact limit of QP energies. However, a quantitative comparison of different levels of theory is worthless if the numerical procedure is not sufficiently reliable.

Numerical issues arise in particular from the unfavorable computational scaling of the $GW$ method, and the slow convergence of the correlation part of the self-energy with respect to the number of unoccupied states makes it difficult to obtain reliable results. This slow convergence stems from the short-range electron correlation and is related to similar issues in quantum-chemistry methods highlighted in Ref.~\onlinecite{Klopper1999}. Yet, there are just a few works~\cite{Friedrich2011,Klimes2014a,Jiang2016} where this fact was explicitly taken into account in order to approach the numerically exact limit of quasiparticle energies or their differences. Several methods to circumvent the slow convergence with respect to the number of unoccupied states have been developed.\cite{Bruneval2008,Berger2010,Umari2011,Lambert2013,Govoni2015} All of them introduce new approximations and numerical algorithms, which should be validated on a large test set and confronted to each other.

A recent paper by Klime\v{s} \textit{et al.}~\cite{Klimes2014a} contains reference-quality $G_0W_0$ data for a range of semiconductors. However, their calculation relies on the projector-augmented waves (PAW) method~\cite{Bloechl1994} which introduces additional numerical uncertainties due to pseudization. This problem is absent in the all-electron full-potential linearized-augmented planewaves (FLAPW) method, which is regarded as the reference method in condensed-matter physics. Friedrich and coworkers have demonstrated an extrapolation procedure for estimating the exact $G_0W_0$ limit for the band gap in wurtzite ZnO based on the FLAPW method,~\cite{Friedrich2011,Friedrich2011a} but they have not applied it to other materials. During the preparation of our manuscript, Jiang and Blaha have published an article~\cite{Jiang2016} in the spirit of the latter work.

In this paper, we show how to construct a basis suitable for excited-state FLAPW calculations and conduct systematic studies on how to reach convergence of quasiparticle energies with respect to the basis size. We provide an interpretation of the exceptional importance of local orbitals in QP calculations utilizing the FLAPW basis and show $G_0W_0$ quasiparticle energies of benchmark quality for a set of materials. Binding energies of semi-core $d$-states are also reported. All calculations are carried out by the \exciting\ code.

The paper is organized as follows. The $G_0W_0$ method is sketched in Sec.~\ref{sec:gw}. Computational details regarding the all-electron basis used in this work and our $G_0W_0$ implementation are described in Sec.~\ref{sec:flapw}. The convergence of the QP gap in wurtzite ZnO, that is considered as a most challenging material, is studied in Sec.~\ref{sec:ZnO-tests}. We reexamine the extrapolation technique suggested by Friedrich\cite{Friedrich2011} and propose an alternative procedure for estimating the exact $G_0W_0$ limit of QP energies. In Sec.~\ref{sec:bandgaps}, we present $G_0W_0$ QP energies for 16 semiconductors and wide-gap insulators. We confront results based on {\it default} and {\it accurate} basis sets. Our results are finally compared to corresponding values of Ref.~\onlinecite{Klimes2014a} obtained within the PAW method.

\section{\label{sec:comp}Methodology}

\subsection{\label{sec:gw}$G_0W_0$ approximation}

The quasiparticle energies within the $G_0W_0$ approximation are given as a solution of the linearized QP equation\cite{Hybertsen1986} as
\begin{align}\label{eq:gw:eqp}
\epsilon_{n\mathbf{k}}^{\rm QP} = \epsilon_{n\mathbf{k}} +
Z_{n\mathbf{k}} 
\left[ 
\Re \; \Sigma_{n\mathbf{k}}(\epsilon_{n\mathbf{k}}) - V^{\rm xc}_{n\mathbf{k}}
\right]\;,
\end{align}
where $\epsilon_{n\mathbf{k}}$ are the Kohn-Sham (KS) eigenvalues, $\Sigma_{n\mathbf{k}}$ and $V^{\rm xc}_{n\mathbf{k}}$ are the diagonal matrix elements of the self-energy and the exchange-correlation (xc) potential that is employed in the single-particle KS Hamiltonian.
The QP renormalization factor $Z_{n\mathbf{k}}$ accounts for the energy dependence of the self-energy.
The non-local and energy-dependent self-energy $\Sigma(\mathbf{r},\mathbf{r}';\omega)$ is given by
\begin{equation}\label{eq:gw:selfe}
\Sigma(\mathbf{r},\mathbf{r}';\omega) = \frac{i}{2\pi} \int\!
G_0(\mathbf{r},\mathbf{r}';\omega+\omega')\: W_0(\mathbf{r},\mathbf{r}';\omega')
\:e^{i\omega'\eta}\:d\omega'\,,
\end{equation}
where $G_0$ is the non-interacting single-particle Green function obtained from the Kohn-Sham states
\begin{equation}\label{eq:gw:G0}
G_0(\mathbf{r},\mathbf{r}';\omega) = \sum_{n\mathbf{k}}
\frac{\psi_{n\mathbf{k}}(\mathbf{r})\:\psi_{n\mathbf{k}}(\mathbf{r}')}
{\omega-\tilde{\epsilon}_{n\mathbf{k}}}\:,
\end{equation}
with $\tilde{\epsilon}_{n\mathbf{k}} \equiv \epsilon_{n\mathbf{k}}+i\eta{~}\mathrm{sgn}(\epsilon_{\rm F}-\epsilon_{n\mathbf{k}})$ and $\eta$ being an infinitesimal positive number.
The dynamically screened Coulomb potential $W_0(\mathbf{r},\mathbf{r}';\omega)$ is determined by
\begin{equation}\label{eq:gw:W0}
W_{0}(\mathbf{r},\mathbf{r}';\omega) = \int\!
\varepsilon^{-1}(\mathbf{r},\mathbf{r}_1;\omega)\:v_{\rm C}(\mathbf{r}_1,\mathbf{r}')\:d\mathbf{r}_1\,,
\end{equation}
where $v_{\rm C}(\mathbf{r},\mathbf{r}')=1/|\mathbf{r}-\mathbf{r}'|$ is the bare Coulomb potential and $\varepsilon(\mathbf{r},\mathbf{r}';\omega)$ is the dielectric function calculated as
\begin{equation}\label{eq:gw:df}
\varepsilon(\mathbf{r},\mathbf{r}';\omega) = \delta(\mathbf{r},\mathbf{r}')-
\int\! v_C(\mathbf{r},\mathbf{r}_1)\: P_0(\mathbf{r}_1,\mathbf{r}';\omega)\: d\mathbf{r}_1\,.
\end{equation}
$P_0(\mathbf{r},\mathbf{r}^\prime;\omega)$ is the irreducible polarizability evaluated in the random-phase approximation
\begin{align}\label{eq:gw:p0}
P_{0}\left(\mathbf{r},\mathbf{r'};\omega \right ) = 
 &
\sum_{n\mathbf{k}} \sum_{m\mathbf{k}'}
F_{nm}(\mathbf{k},\mathbf{k}';\omega) \times \\ \nonumber
 & 
\psi_{n\mathbf{k}}(\mathbf{r}) \psi_{m\mathbf{k}'}^{*}(\mathbf{r})
\psi_{n\mathbf{k}}^{*}(\mathbf{r}')\psi_{m\mathbf{k}'}(\mathbf{r}')
\end{align}
with
\begin{align}\label{eq:gw:fnm}
 & F_{nm}(\mathbf{k},\mathbf{k}';\omega) \equiv 
2f_{n\mathbf{k}} [1-f_{m\mathbf{k}'}] \\\nonumber &
\left\{\frac{1}{\omega-\epsilon_{m\mathbf{k}'}+\epsilon_{n\mathbf{k}}+i\eta} -
\frac{1}{\omega+\epsilon_{m\mathbf{k}'}-\epsilon_{n\mathbf{k}}-i\eta} \right\},
\end{align}
where $f_{n\mathbf{k}}$ is the occupation number of the state $n\mathbf{k}$ and the factor of 2 accounts for spin degeneracy.

\subsection{\label{sec:flapw}LAPW and local orbitals}
Our calculations employ the linearized augmented planewaves + local orbitals (LAPW+lo) basis~\cite{Singh2006} as implemented in the \exciting{} code.\cite{Gulans2014a} In this method, the unit cell is partitioned into the muffin-tin (\textit{MT}) and interstitial (\textit{I}) regions.
The LAPW part of the basis is determined as
\begin{equation}
\label{eq:basis}
\phi^{\mathbf{k}}_\mathbf{G}(\mathbf{r})= \left\{
\begin{array}{cl}
  \sum\limits_{lm\zeta} A^\mathbf{G+k}_{lm\alpha\zeta} u_{l\alpha\zeta} (r_\alpha,\epsilon_{l\alpha}) Y_{lm}(\hat{\mathbf{r}}_\alpha) &  
  \mathbf{r}\in MT_{\alpha}\\
  \frac{1}{\sqrt{\Omega}} e^{i(\mathbf{G+k})\mathbf{r}} & \mathbf{r}\in I
\end{array}
\right.
\end{equation}
where $MT_\alpha$ refers to a sphere with radius $R_\mathrm{MT}$ that is centered at the position $\mathbf{R}_\alpha$ ($\mathbf{r}_\alpha=\mathbf{r}-\mathbf{R}_\alpha$) of atom $\alpha$. The $MT$ part of $\phi^{\mathbf{k}}_\mathbf{G}(\mathbf{r})$ is expanded in terms of spherical harmonics $Y_{lm} (\hat{\mathbf{r}}_\alpha)$ and radial functions $u_{l\alpha\zeta}(r_\alpha,\epsilon_{l\alpha})$. The latter are solutions of the radial scalar-relativistic Schr\"{o}dinger equation at fixed reference energies $\epsilon_{l\alpha}$ (linearization energies) inside each $MT$ employing a spherically averaged Kohn-Sham effective potential. The index $\zeta$ accounts for different types of radial functions, which are either $u_{l\alpha}(r_\alpha,\epsilon_{l\alpha})$ or $\dot{u}_{l\alpha}(r_\alpha,\epsilon_{l\alpha}) \equiv \left. \frac{\partial u_{l\alpha}(r_\alpha,E)}{\partial E} \right|_{\epsilon_{l\alpha}} $. The augmentation coefficients $A^\mathbf{G+k}_{lm\alpha\zeta}$ ($\zeta$=1,2 in the LAPW method) are chosen to make $\phi^{\mathbf{k}}_\mathbf{G}(\mathbf{r})$ continuous and smooth at the sphere boundary.

The LAPW basis set is complemented with functions $\phi_\mu^{lo}(\mathbf{r})$ \cite{Singh1991,Sjostedt2000}, local orbitals, which are strictly zero everywhere, except for a certain $MT$ sphere:
\begin{equation}
\label{eq:lo}
\phi_\mu^{lo}(\mathbf{r})= \left[ a_\mu f_\mu(r_{\alpha})+b_\mu g_\mu(r_{\alpha}) \right] \;Y_{lm}(\hat{\mathbf{r}}_\alpha).
\end{equation}
Generally, $f_\mu(r_{\alpha})$ and $g_\mu(r_{\alpha})$ can be any smooth radial functions, but in the present work, we employ $u_{l\alpha}(r_\alpha,\epsilon_{l\alpha})$ and their energy derivatives. The coefficients $a_\mu$ and $b_\mu$ are determined to turn the local orbital to zero at the $MT$ boundary and to fulfill the normalization condition $\langle\phi_\mu^{lo} | \phi_\mu^{lo}\rangle=1$.

In the present work, we construct two setups of local orbitals for each element in every system.
The first one contains a small number of lo's, suitable for ground state calculations (we will refer to it as ``{\it default}'' thereafter). For example, for computing ZnO, it contains 4 $s$-, 4 $p$- and 2 $d$-shells of local orbitals for the zink atom and 2 $s$- and 2 $p$-shells for oxygen. The second setup (labeled as ``{\it optimized}'') is constructed with the aim of very accurately describing unoccupied states. In practical calculations, we use local orbitals with $l$ up to 8 and $\epsilon_{l}^\mathrm{lo}<100$~Ha. With these limits, Zn and O atoms with $MT$ radii of 1.6~$a_0$ are described with 6 $s$-, 7 $p$-, 7 $d$-, 7 $f$-, 6 $g$-, 6 $h$-, 5 $i$-, 5 $j$- and 5 $k$-shells of local orbitals. More details on the construction of local orbitals are given in the Appendix~\ref{app:losetups}. For comparison, the setup used in Ref.~\onlinecite{Friedrich2011} consists of 5 $s$-, 5 $p$-, 5 $d$-, 5 $f$-, 3 $g$-, 2 $h$- and 1 $i$-shells for Zn and 4 $s$-, 4 $p$-, 4 $d$-, 4 $f$-, 2 $g$- and 1 $h$-shells for O.

\subsection{\label{sec:implementation}Implementation}

Our $G_0W_0$ implementation employs an auxiliary mixed product basis\cite{Kotani2002,Friedrich2010,Jiang2013} (MB) which provides a highly flexible representation of the polarizability, the dielectric function, and other non-local quantities.
Moreover, the usage of the MB allows us to convert the expressions Eqs.~\eqref{eq:gw:selfe}--\eqref{eq:gw:df} into a computationally tractable matrix form.
Due to the dual nature of the LAPW+lo basis, the MB consists of spherical harmonics and planewaves within $MT$'s and $I$, respectively.
An optimal set of these MB functions, $\chi_{i}^{\mathbf{q}}(\mathbf{r})$, accurately represents a product of two KS wave functions such as 
\begin{equation}
  \phi_{n\mathbf{k}}(\mathbf{r})\phi_{m\mathbf{k}-\mathbf{q}}^{*}(\mathbf{r}) = \sum_{i} M^{i}_{nm}(\mathbf{k},\mathbf{q}) \chi_{i}^{\mathbf{q}}(\mathbf{r}),
\end{equation}
where, $ M^{i}_{nm}(\mathbf{k},\mathbf{q})$ are the expansion coefficients and $\mathbf{q} = \mathbf{k}-\mathbf{k}'$.
The quality of the mixed basis, in analogy to the LAPW+lo basis, is controlled by setting the maximal angular momentum for the spherical-harmonics expansion, $L_{MB}$, and the number of planewaves according to the cutoff parameter $G_{MB}$.
To exclude any effects connected with the representation of the non-local operators, we have chosen rather high values of $L_{MB}=12$ and $G_{MB}=G_{max}^{LAPW}$ in our calculations.

The number of the empty states, the size of $\mathbf{k}$ and $\mathbf{q}$ grids, and the number of frequencies used to represent the correlation self-energy and to calculate the convolution integral (Eq.~\ref{eq:gw:selfe}) are the most important computational parameters.
According to previous studies,\cite{Friedrich2011,Klimes2014a} the $\mathbf{k}/\mathbf{q}$-grid convergence is found to be independent of the convergence w.~r.~t. the number of the empty states.
Based on our convergence tests (see Appendix~\ref{app:kqtest}, for the purpose of computational efficiency, we perform the $G_0W_0$ calculations for a large number of empty states on a coarse grid ({\it e.g.}, $2\times2\times2$ or $4\times4\times4$).
The values for denser $\mathbf{k}/\mathbf{q}$-grids are then extrapolated by applying a constant shift that is devised from $\mathbf{k}/\mathbf{q}$-grid convergence tests for a small number of empty states (typically 200).
We find that results obtained with this procedure and a direct calculation employing a dense grid agree within 50 meV.

Polarizability, screened Coulomb potential, and correlation self-energy are calculated on a non-uniform grid of 32 imaginary frequencies.
As the final step, the QP energies (Eq.~\eqref{eq:gw:eqp}) are calculated employing the Pad\'e-approximant method\cite{Vidberg1977} for the analytical continuation.
Further technical details on our implementation can be found in Refs.~\onlinecite{Jiang2013,Gulans2014a}.

\section{\label{sec:tests}Results}

\subsection{\label{sec:ZnO-tests}Quasiparticle gap of wurtzite Z\MakeLowercase{n}O}

\subsubsection{\label{sec:extrapol}Extrapolation to infinite number of states}

The calculation of $\epsilon_{n\mathbf{k}}^{\rm QP}$ is a computationally demanding process, as it involves two summations over states. These sums appear in Eqs.~\eqref{eq:gw:G0} and \eqref{eq:gw:p0} and formally include an infinite number of unoccupied bands. The common practice, however, is to consider a relatively small number of unoccupied bands. Such a simplification makes $G_0W_0$ affordable for a wide range of applications. However, the truncation inevitably introduces an error as it asymptotically decays as 
\begin{equation}
\label{eq:asymp}
\Delta \epsilon^{\rm QP}\sim 1/N,
\end{equation}
where $N$ is the number of unoccupied bands. The slow decay originates from a two-electron coalescence; see Refs.~\onlinecite{Schindlmayr2013, Gulans2014, Klimes2014a} for a detailed discussion.

As the behavior of $\Delta \epsilon^{\rm QP}$ is known, one naturally attempts to estimate the limit of the quasiparticle-energy by extrapolating $\epsilon^{\mathrm{QP}}_{\mathrm{n\mathbf{k}}}(N)$ to infinite $N$. Such an approach was used by Friedrich and coworkers in Ref.~\onlinecite{Friedrich2011}. Their extrapolation procedure required a calculation of QP gaps $E^{\mathrm{QP}}_{\mathrm{g}}(N)$ for a range of $N$. The obtained data were then fitted to the expression
\begin{align}
\label{eq:extrapol:gw}
E^{\mathrm{QP}}_{\mathrm{g}}(N) = E^{\mathrm{QP}}_{\mathrm{g}}(\infty)+\frac{a}{(b+N)},
\end{align}
where $a$, $b$, and $E^{\mathrm{QP}}_{\mathrm{g}}(\infty)$ are fit parameters.
Eq.~\eqref{eq:extrapol:gw} is consistent with Eq.~\eqref{eq:asymp}, and we adopt this procedure in this section.

\begin{figure}[tbp] 
\begin{center}
\includegraphics[width=0.9\columnwidth]{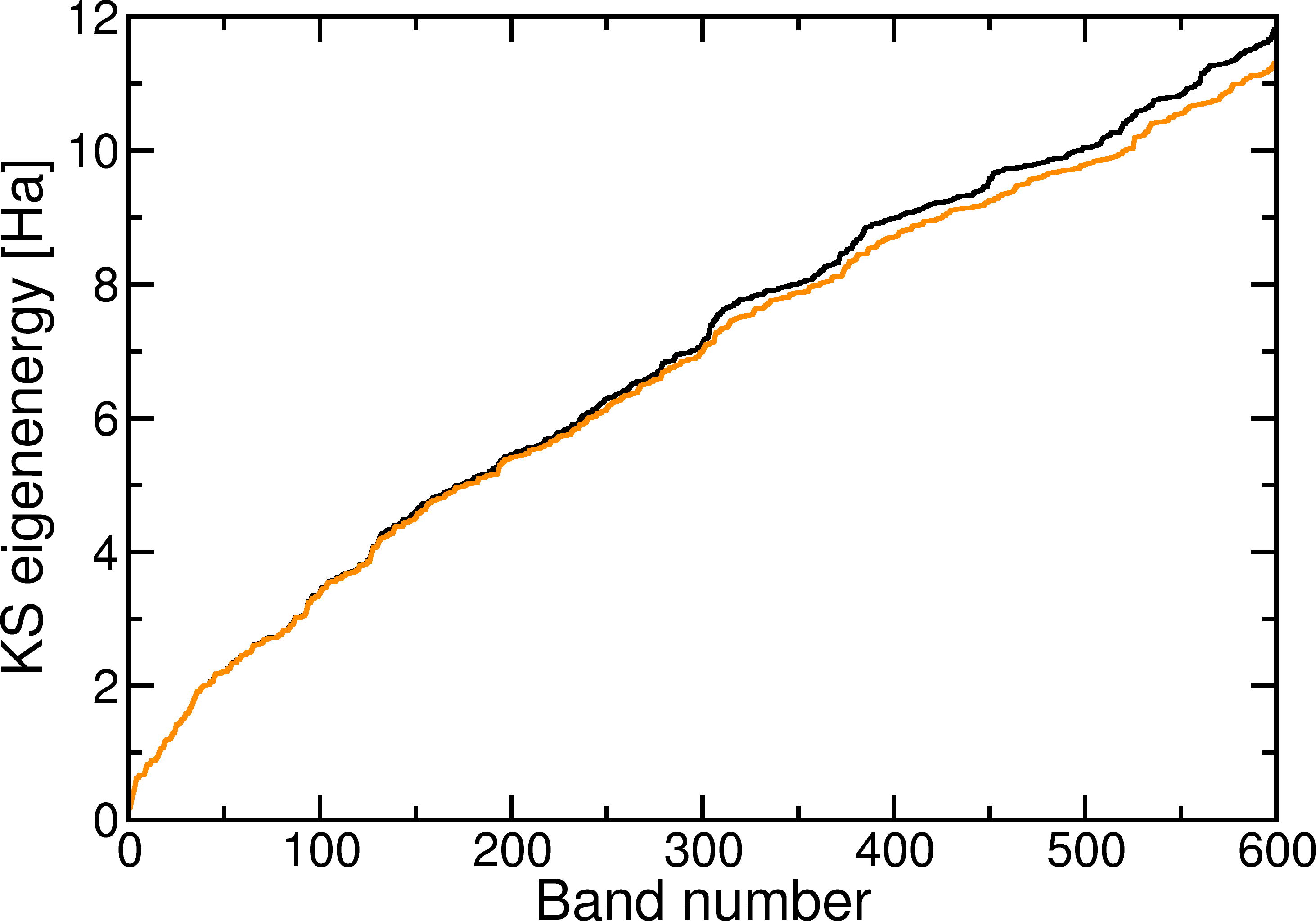}
\end{center}
\caption{\label{fig:ks-lapw-vs-lapwlo}
(Color online) Wurtzite ZnO KS eigenenergies at the $\Gamma$-point.
The black curve corresponds to a calculation with the set of local orbitals defined in Table~\ref{tab:lo}, while the red one is obtained combining the latter basis with local orbitals for unoccupied states as defined in Table~\ref{tab:xslo}.}
\end{figure}

\begin{figure}[htbp]
\begin{center}
\includegraphics[width=\columnwidth]{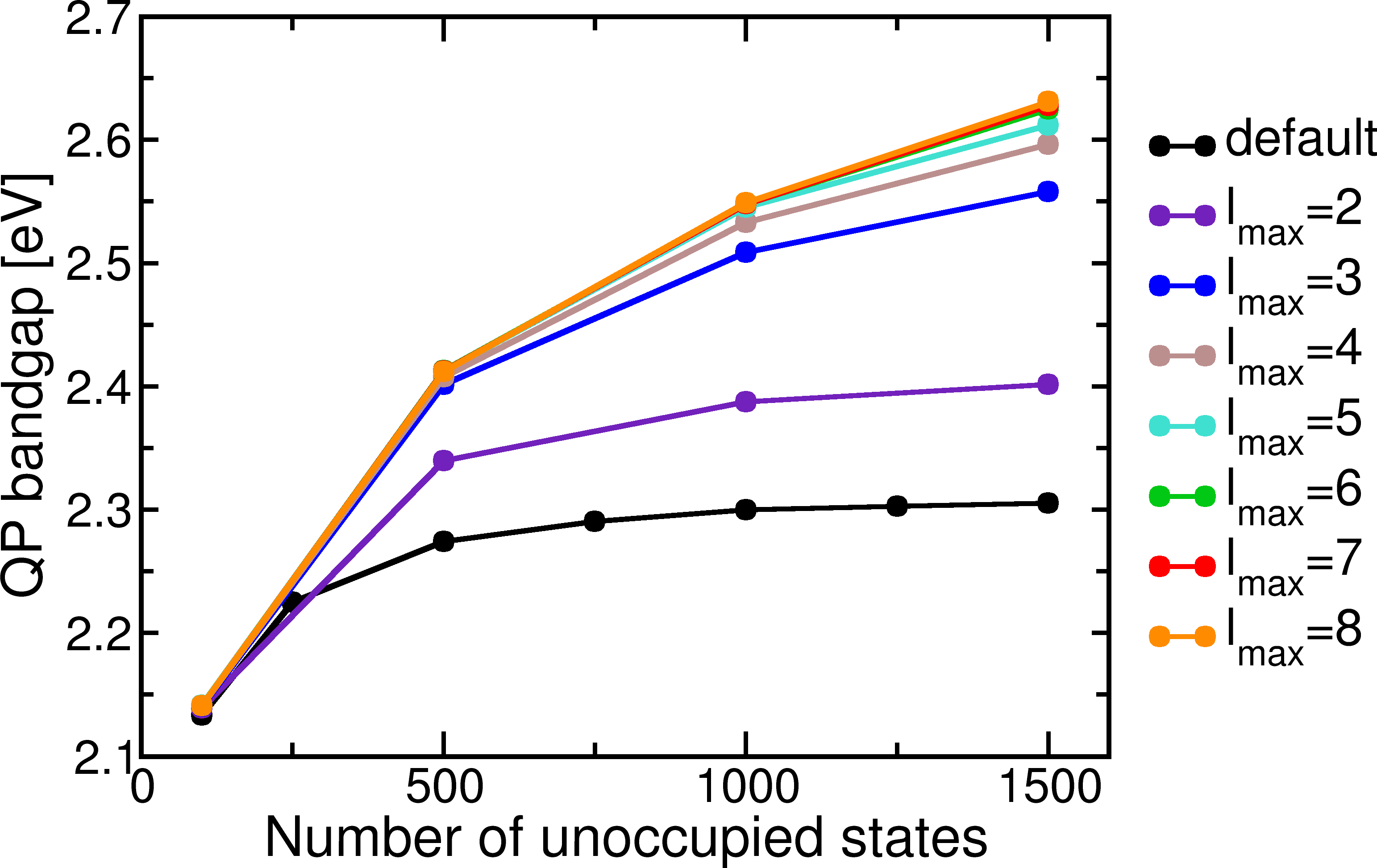}
\end{center}
\caption{\label{fig:qpgap-lo-sets}
Convergence of the QP bandgap of wurtzite ZnO for different configurations of the basis set.
Each configuration is characterized by a maximum angular momentum $l_\mathrm{max}$ up to which additional local orbitals for representing unoccupied states are added (see Table~\ref{tab:xslo}).
}
\end{figure}

It was also shown in Ref.~\onlinecite{Friedrich2011} that quasiparticle calculations follow Eqs.~\eqref{eq:asymp} and \eqref{eq:extrapol:gw} only if the unoccupied states in the considered range are described accurately enough. Thus, it is crucially important to employ a sufficiently flexible basis. In order to do so, we expand the basis as described in Sec.~\ref{sec:flapw} and Appendix~\ref{app:losetups}. In this work, all groundstate calculations are performed in the local-density approximation (LDA).~\cite{Perdew1992} The resulting KS spectrum is shown in Fig.~\ref{fig:ks-lapw-vs-lapwlo} and compared to the that of the {\it default} settings. While the setups designed for the ground state on the one hand, and unoccupied states on the other hand produce nearly identical KS energies for the first hundred states, the difference is substantial for high-lying states. Although, the two curves look qualitatively similar, both behaving roughly as $\epsilon(N)\sim\sqrt{N}$, the energies of the 2000th unoccupied state differ by $\sim 0.8$~Ha ($\sim 3\%$).
Despite the seemingly small impact on the KS spectrum, the additional local orbitals introduce a dramatic change in the QP gap of ZnO as shown in Fig.~\ref{fig:qpgap-lo-sets}. (Note that all $G_0W_0$ calculations in this section have been performed on coarse $2\times2\times2$ \textbf{k}/\textbf{q}-grids, and the figure does not reflect the final higly-converged values). The {\it default} setup (Table~\ref{tab:lo}) leads to a gap that saturates to $\sim 2.3$~eV at a high number of unoccupied bands.

To examine the basis-set quality, we perform $G_0W_0$ calculations with different basis sets that are characterized by the maximum angular momentum, $l_\mathrm{max}$, up to which local orbitals have been added. All details regarding their configurations are specified in Table~\ref{tab:xslo}. As it follows from Fig.~\ref{fig:qpgap-lo-sets}, the extensions of the {\it default} configuration produce notable changes in the bahavior of the QP gap. Employing the largest basis set with $N = 1300$ empty bands, the gap reaches the value of 2.6 eV but still keeps growing without any sign of saturation. A qualitatively similar behavior was previously reported by Friedrich in Refs.~\onlinecite {Friedrich2011, Friedrich2011a}.

Analyzing the behavior of $E^{\mathrm{QP}}_{\mathrm{g}}(N)$ as a function of $l_\mathrm{max}$, one can immediately state the exceptional importance of the basis quality for obtaining reliable KS states. Thus, according to Fig.~\ref{fig:qpgap-lo-sets}, for the treatment of unoccupied states, additional basis functions (local orbitals) must be added into all angular momentum channels up to $l_\mathrm{max} = 6$. As it was shown earlier,\cite{Friedrich2011} the basis-set incompleteness error is responsible for reaching a spurious convergence of the bandgap in wurtzite ZnO for a moderate number of empty states.

Following all these findings, we generate new basis sets and apply the extrapolation procedure to all studied compounds.Technically, we use a set of empty states, typically consisting of independent calculations in the relevant interval concerning the number of unoccupied states as determined by $G_\mathrm{max}$. The corresponding results will be presented in Section~\ref{sec:bandgaps}.

\begin{figure}[htbp]
\begin{center}
\includegraphics[width=0.8\columnwidth]{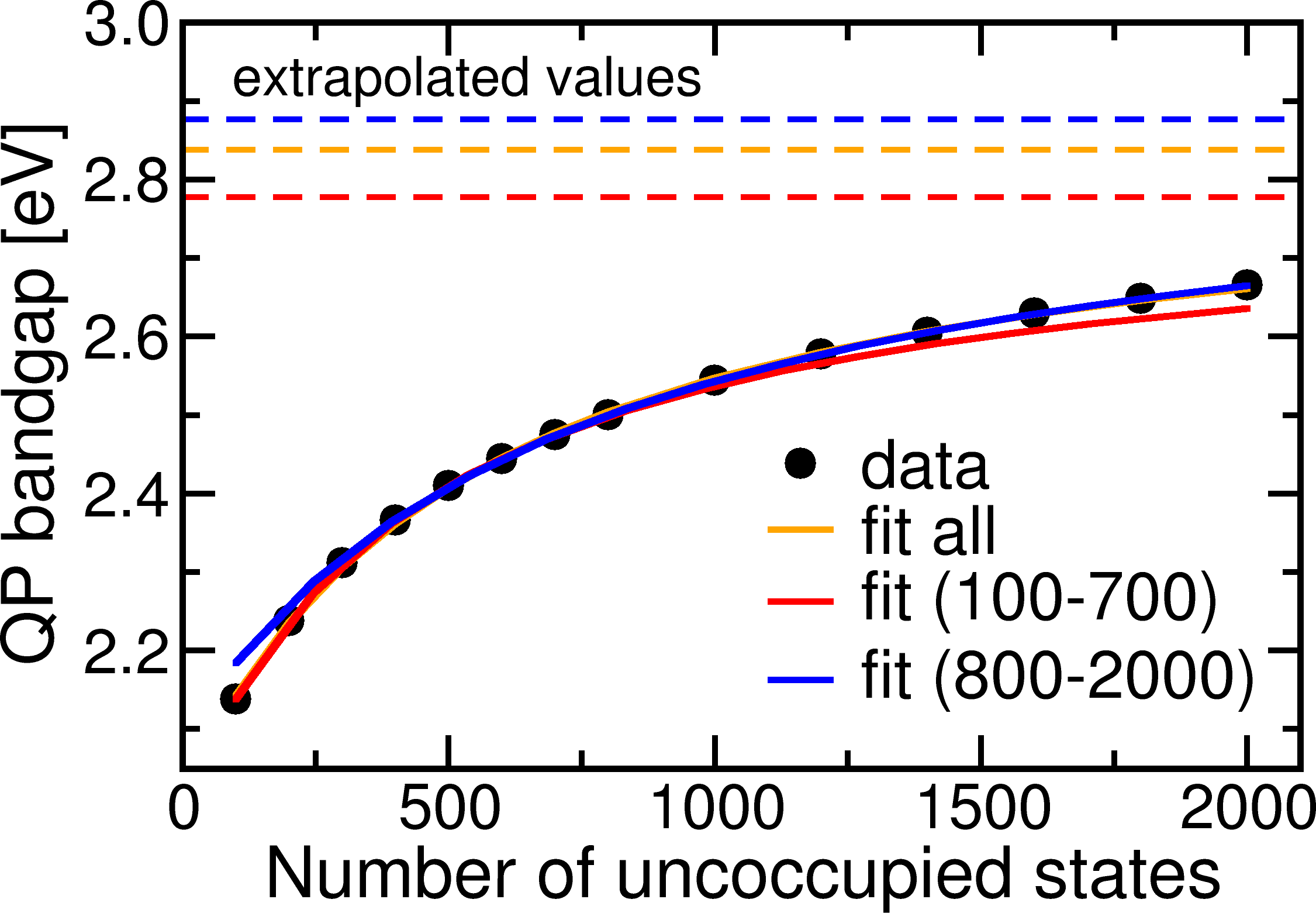}
\end{center}
\caption{\label{fig:ZnO-extrapolation}
(Color online) Uncertainty of the extrapolation procedure with respect to the interval of empty bands (specified in parentheses) used for fitting Eq.~\eqref{eq:extrapol:gw}.
The dashed lines on top represent the extrapolated values.}
\end{figure}

Although this extrapolation procedure is straightforward, it has a major practical drawback. This concerns the question, which range of unoccupied states $N$ is suitable for the fitting. First, high-energy unoccupied KS states may still contain an error that impacts the extrapolated gap (cf. Fig.~\ref{fig:ks-lapw-vs-lapwlo}). From this point of view, it is undesirable to include results obtained with too large $N$ in the extrapolation procedure, i.e. states that cannot be fully trusted. Second, it is not \textit{a priori} obvious at which $N$ the QP gap starts to follow the asymptotic behavior described by Eq.~\ref{eq:extrapol:gw}. Applied to wurtzite ZnO, this procedure yields  extrapolated gaps in the range of 2.79--2.89~eV, depending on the fitting range $N$ as shown in Fig.~\ref{fig:ZnO-extrapolation}. The lower (higher) limit is obtained when we use $N=100-700$ ($N=1100-2000$) for the fit.

\subsubsection{\label{sec:all}All-states calculations}

Since the extrapolation to an infinite number of unoccupied states introduces an uncertainty, as just dicsussed above, we introduce an alternative method for obtaining converged QP energies.
This method involves the entire KS spectrum that can be obtained with a finite yet sufficiently rich LAPW+lo basis set.
It is based on our observation that, while it is important to introduce a relatively large number of local orbitals, the quasiparticle energies are only weakly influenced by the number of LAPW's.
As a consequence, such an {\it all-states} calculation is still feasible, while it yields the limit of an infinite number of states without any extrapolation.

\begin{figure}[htbp] 
\begin{center}
\includegraphics[width=0.9\columnwidth]{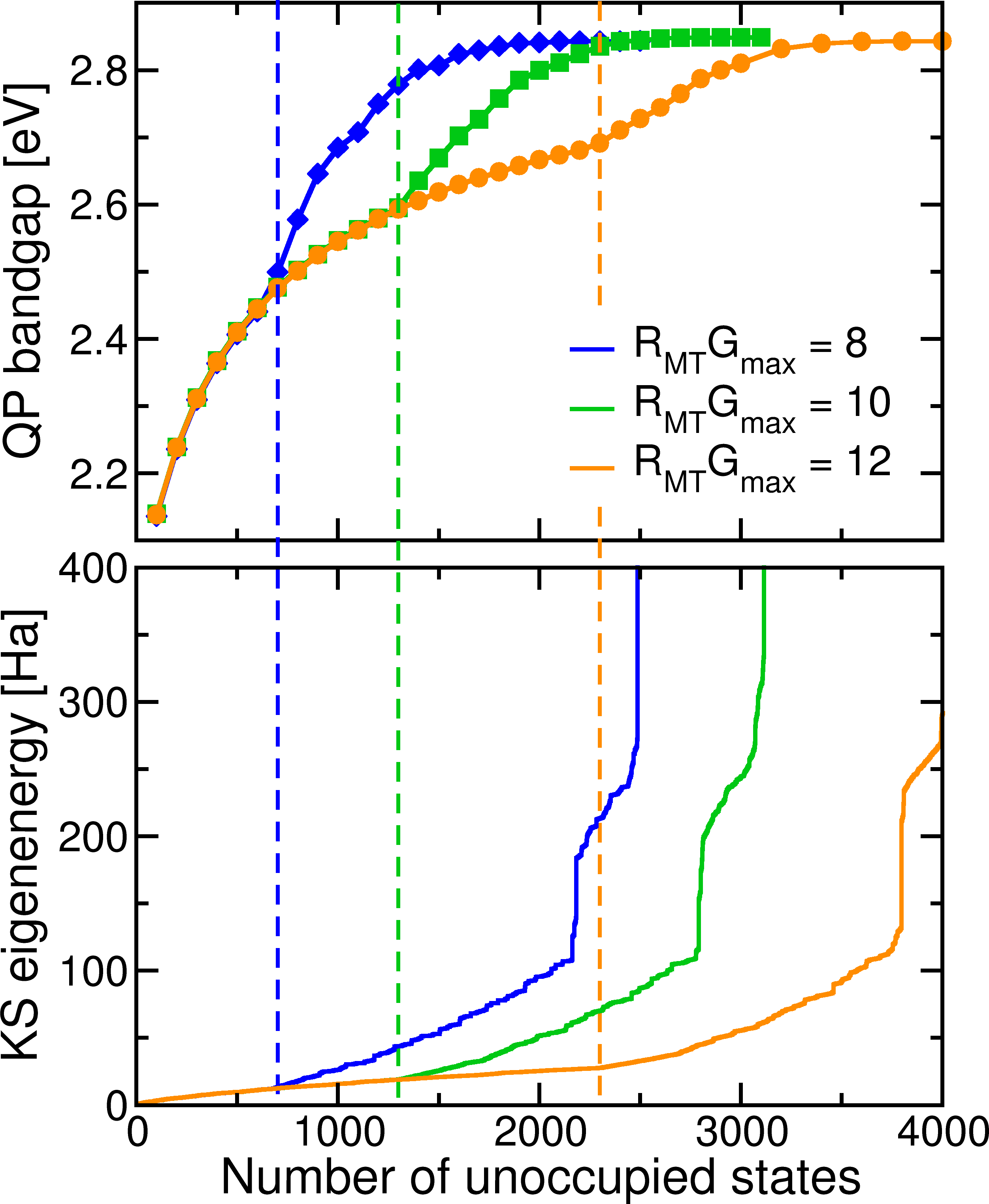}
\end{center}
\caption{\label{fig:all-states}
(Color online) Top: $G_0W_0$ gap as a function of the number of unoccupied bands used in the calculations.
The blue, green, and orange curves correspond to calculations with $R_\mathrm{MT}G_\mathrm{max}$ values of 8, 10 and 12, respectively.
In all cases, local orbitals up to $l_{\textrm{max}}=8$ were employed.
Bottom: Corresponding KS energies of unoccupied bands at the $\Gamma$-point.
Vertical dashed lines indicate the threshold values, $N_0$, at which $E^\mathrm{QP}_\mathrm{g}(N)$ stops following the asymptotic behavior Eq.~\ref{eq:asymp}.}
\end{figure}

In order to justify such an approach, we reexamine the convergence behavior of $E^\mathrm{QP}_\mathrm{g}(N)$ taking into account the dual nature of LAPW+lo basis set.
At first, the local basis size is fixed and consists of local orbitals with $l\leq 8$, while the number of LAPW's is varied.
As displayed in Fig.~\ref{fig:all-states} (top), $E^\mathrm{QP}_\mathrm{g}(N)$ behaves according to Eq.~\ref{eq:asymp} roughly until the threshold $N=N_0$ (dashed vertical lines) that depends on the total number of LAPW functions in the basis (as determined by $R_{\mathrm{MT}}G_{\mathrm{max}}$).
At $N>N_0$, $E^\mathrm{QP}_\mathrm{g}(N)$ deviates from the hyperbola and saturates as $N$ approaches its maximum.
Such a threshold ($N_0$) exists, because an accurate description of high-energy unoccupied states requires LAPW's with larger $|\mathbf{G+k}|$, however, their number is limited in practice.
In other words, KS orbitals in the region $N>N_0$ are described mostly by local orbitals.
This conclusion becomes apparent in Fig.
~\ref{fig:all-states} (bottom), where the KS energies $\epsilon_{n\mathbf{k}}(N)$ at the $\Gamma$ point are plotted.
They approximately follow the usual $\sqrt{N}$-rule up to the threshold that coincides with $N_0$ in Fig.~\ref{fig:all-states} (top).
At $N>N_0$, $\epsilon_{n\mathbf{k}}(N)$ grows much more rapidly than $\sqrt{N}$.

Fig.~\ref{fig:all-states} (top) also reveals a striking detail. All three curves which are obtained by using different numbers of LAPW's saturate at the same value of 2.84~eV within 5 meV. Thus, LAPW's, in contrast to local orbitals, have a weak impact on the convergence of quasiparticle gaps and, more generally, quasiparticle energies. In other words, it is much more important to have a rich basis within the MT's than in the interstitial region.

We can understand this by the following considerations. If all unoccupied bands are used in a quasiparticle calculation, all degrees of freedom provided by this LAPW+lo basis contribute to $P_0(\mathbf{r},\mathbf{r}^\prime,\omega)$. Extending the KS basis improves the description of $P_0(\mathbf{r},\mathbf{r}^\prime,\omega)$, which is difficult to resolve in the limit of short $\mathbf{r}^\prime - \mathbf{r}$. As shown is Ref.~\onlinecite{Gulans2014}, a poor description of short-range features of $P_0(\mathbf{r},\mathbf{r}^\prime,\omega)$ and related quantities has the strongest effect on those states that overlap substantially with high-density regions. Based on this observation, we argue that, in order to improve quasiparticle energies, one has to expand the KS basis in high-density regions, thus, better resolving the short-range features of $P_0(\mathbf{r},\mathbf{r}^\prime,\omega)$ there. Since MT's enclose high-density regions, adding local-orbitals has a noticeable effect on the quasiparticle energies, while increasing the number of LAPW's has a minor influence as stated above.

\begin{figure}[htbp] 
\begin{center}
\includegraphics[width=0.9\columnwidth]{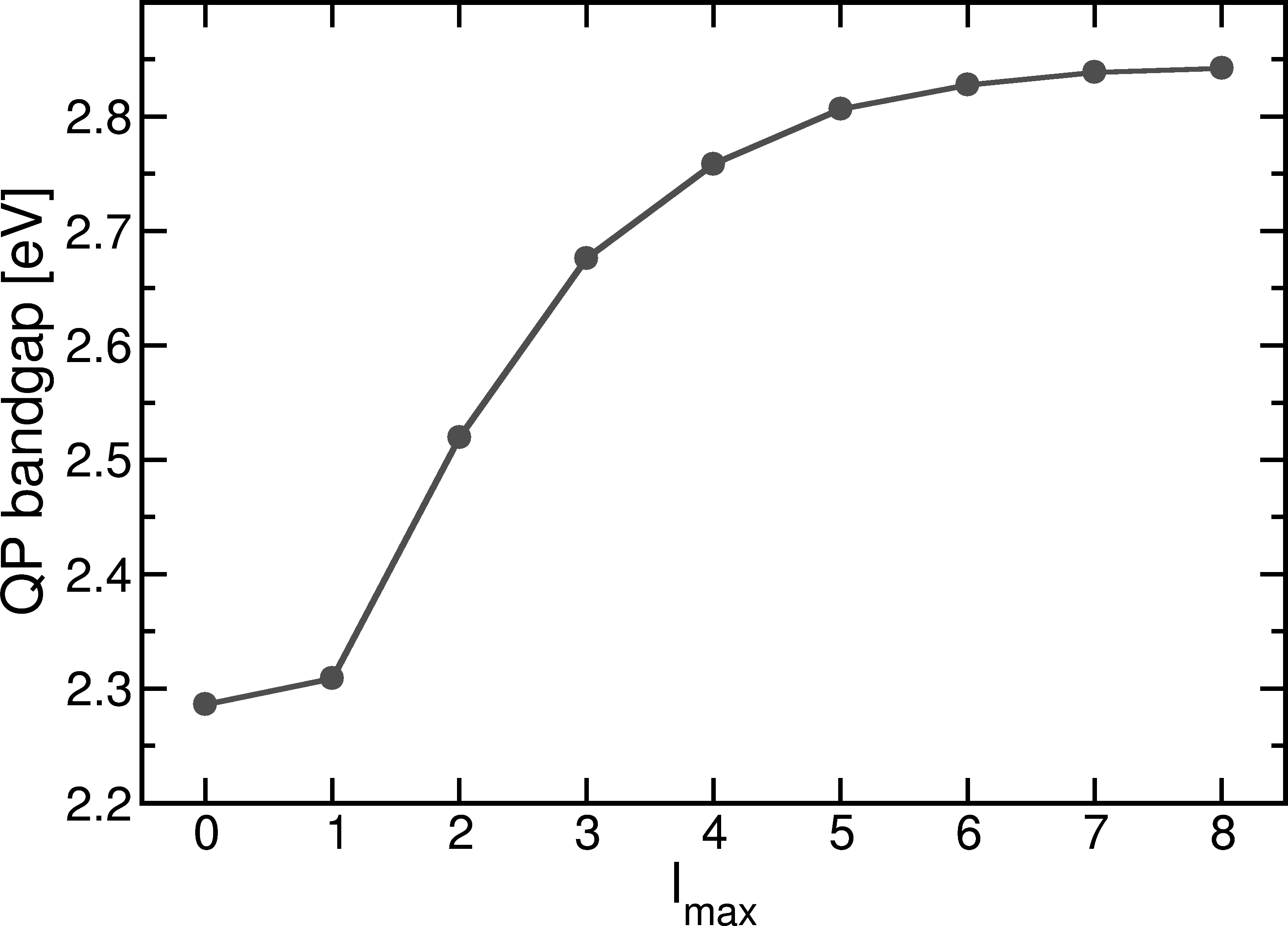}
\end{center}
\caption{\label{fig:gw-lmax}
(Color online) $G_0W_0$ gap as a function of $l_\mathrm{max}$ as used to construct additional local orbitals for unoccupied states.}
\end{figure}

Further, we explore how well converged the calculations with our local-orbital settings are.
Fig.~\ref{fig:gw-lmax} shows how $l_\mathrm{max}$ influences the quasiparticle gap when using all available unoccupied states. Setups with $l_\mathrm{max}=0$ and $1$ lead to a value of around 2.3~eV that coincides with the large-$N$ limit for the {\it default} setup. For $l_\mathrm{max}>1$, the quasiparticle gap initially increases rapidly, but then it saturates at around $l_\mathrm{max}=7-8$. The difference between $l_\mathrm{max}=7$ and 8 is less than 5 meV.

While the angular degrees of freedom spanned by the local-orbital basis have a significant impact on the quasiparticle energies, the influence of the radial degrees of freedom is not to be overlooked.
To test whether the latter ones are sufficiently well represented, we have performed a calculation with $l_\mathrm{max}=8$ and an energy-cutoff of $160$~Ha. These settings correspond to 8 $s$-, 9 $p$-, 9 $d$-, 9 $f$-, 8 $g$-, 8 $h$-, 7 $i$-, 7 $j$-, and 6 $k$-shells of local orbitals for both Zn and O atoms (2404 local orbitals in total). The quasiparticle gap increases by 3~meV as the result of the basis adjustments. Thus, we conclude that both the angular and radial degrees of freedom are sufficiently well represented by the basis described in Sec.~\ref{sec:flapw} and Appendix~\ref{app:losetups}.

\subsection{\label{sec:bandgaps}$G_0W_0$ bandgaps for semiconductors and insulators}

We have calculated the quasiparticle bandstructure for a test set of compounds consisting of 13 small- and wide-gap semiconductors as well as 3 insulators. The corresponding structures and lattice parameters have been taken from Refs.~\onlinecite{Gruneis2014,Shishkin2007} and are provided in Appendix~\ref{app:crystals}. The carefully converged groundstate calculations have been performed within LDA. The resulting KS eigenvalues of the highest valence band (VB) and lowest conduction band (CB) at the $\Gamma$, $X$, and $L$ points of the Brillouin zone are presented in Table~\ref{tab:qp-lda}. Comparing these values with those of Ref.~\onlinecite{Klimes2014a}, we find excellent agreement. For the valence region, the differences do not exceed 20 meV. The deviations are somewhat larger for the positions of the $d$-states ($\sim 50$ meV).
\setlength{\tabcolsep}{5pt}
\begin{table}[tbp]
\caption{\label{tab:qp-lda}Kohn-Sham (LDA) energies of the valence band (VB) at $X$ and $L$, and the conduction band (CB) at $\Gamma$, $X$, and $L$ with respect to the VB maximum at $\Gamma$, thus $\Gamma_c$ reflecting the direct band gap.
For Ga, Zn, and Cd the averaged position of $d$-states at $\Gamma$ is also indicated.}
\begin{tabular}{ l r r r r r r}
  \hline\hline
Material	& $\Gamma_c$ & $L_v$ & $L_c$ & $X_v$ & $X_c$ & $\Gamma_d$ \\
\hline
C 	 &  5.55 & -2.79 &  8.39 & -6.29 &  4.71 &  \\
SiC  &  6.28 & -1.06 &  5.38 & -3.19 &  1.32 &  \\
Si   &  2.53 & -1.20 &  1.42 & -2.85 &  0.61 &  \\
BN   &  8.69 & -1.95 & 10.20 & -4.92 &  4.35 &  \\
AlP  &  3.09 & -0.77 &  2.65 & -2.12 &  1.45 &  \\
GaN-zb &  1.62 & -0.96 &  4.41 & -2.66 &  3.25 & -13.45 \\
GaN-wz &  1.93 &       &       &       &       & -13.32 \\
GaAs &  0.30 & -1.15 &  0.85 & -2.69 &  1.35 & -14.86 \\
MgO  &  4.67 & -0.66 &  7.76 & -1.36 &  8.91 &  \\
ZnO-zb  &  0.62 & -0.80 &  5.32 & -2.22 &  5.14 & -5.28 \\
ZnO-wz  &  0.75 &       &       &       &       & -5.22 \\
ZnS  &  1.85 & -0.87 &  3.09 & -2.23 &  3.20 & -6.28 \\
CdS  &  0.88 & -0.78 &  2.77 & -1.95 &  3.30 & -7.59 \\
\hline\hline 
Ar   &  8.18 & -0.15 & 11.05 & -0.45 & 10.85 &  \\
Ne   & 11.44 & -0.08 & 17.05 & -0.23 & 18.34 &  \\  
LiF  &  8.94 & -0.24 & 10.46 & -1.05 & 14.55 &  \\
\hline\hline
\end{tabular} 
\end{table}

Based on these results, the $G_0W_0$ quasiparticle energies have been computed.
Table~\ref{tab:qp-gw} summarizes the data for VB and CB at $\Gamma$, $X$, and $L$, the absolute shifts of the VB maximum compared to LDA ($\Delta$IP), as well as the $d$-band position at $\Gamma$ for Ga, Zn, and Cd.
In all cases, the values obtained for two different basis sets are shown.
The first one (termed {\it optimized}) is generated for each compound in an automatic manner, with local orbitals added to each angular momentum $l \le 8$ and linearization energies in an interval up to 100 Ha.
Employing this basis set, the extrapolated QP energies have been deduced from seven calculations, with empty states ranging from 200 to 800, except for Ar, Ne, wz-GaN and wz-ZnO.
For Ar and Ne, only 5 calculations with 100 to 300 empty states have been used.
In case of wz-GaN and wz-ZnO, the extrapolation has been performed on data obtained with 200--2000 empty states.
To stress the importance of the basis sets for $G_0W_0$ calculations, 
we show for comparison the results obtained with the basis sets that are typically employed in FLAPW groundstate calculations.
As it was shown before in Fig.~\ref{fig:qpgap-lo-sets} for ZnO, the usage of the {\it default} configurations in all cased leads to spurious convergence of $E^{QP}_{g}(N)$.
\setlength{\tabcolsep}{5pt}
\begin{table}[htbp]
\caption{\label{tab:qp-gw} $G_0W_0$ results for the absolute shift of VB compared to LDA ($\Delta$IP) and the positions of VB at $X$ and $L$, and CB at $\Gamma$, $X$, and $L$ with respect to the VB maximum at $\Gamma$; thus $\Gamma_c$ reflecting the direct band gap.
For Ga, Zn, and Cd, the position of the $d$-band at $\Gamma$ is also shown.
The first (second) row presents the results for the {\it optimized} ({\it default}) basis set using the extrapolation procedure.}
\begin{tabular}{l r r r r r r r}
\hline\hline
Material	& $\Delta$IP & $\Gamma_c$ & $L_v$ & $L_c$ & $X_v$ & $X_c$ & $\Gamma_d$ \\
\hline
C 	 & -0.99 &  7.43 & -2.94 & 10.38 & -6.58 &  6.26 &  \\
     & -0.74 &  7.41 & -2.97 & 10.35 & -6.64 &  6.14 &  \\
\hline
SiC  & -0.93 &  7.37 & -1.11 &  6.63 & -3.31 &  2.45 &  \\
     & -0.66 &  7.39 & -1.12 &  6.56 & -3.37 &  2.29 &  \\ 
\hline
Si   & -0.60 &  3.24 & -1.22 &  2.09 & -2.86 &  1.25 &  \\
     & -0.51 &  3.21 & -1.23 &  2.11 & -2.92 &  1.21 &  \\ 
\hline
BN   & -1.47 & 11.28 & -2.04 & 12.32 & -5.13 &  6.47 &  \\
     & -1.12 & 11.10 & -2.09 & 12.16 & -5.24 &  6.23 &  \\ 
\hline
AlP  & -0.81 &  4.10 & -0.78 &  3.69 & -2.14 &  2.41 &  \\
     & -0.52 &  4.14 & -0.80 &  3.65 & -2.20 &  2.25 &  \\  
\hline
GaN-zb & -1.11 &  3.00 & -0.94 &  6.10 & -2.64 &  4.71 & -16.11 \\
       & -0.55 &  2.79 & -1.02 &  5.85 & -2.83 &  4.34 & -15.41 \\
\hline
GaN-wz & -1.09 &  3.35 &       &       &       &       & -15.90 \\
       & -0.64 &  3.20 &       &       &       &       & -15.14 \\
\hline
GaAs & -0.58 &  1.16 & -1.16 &  1.60 & -2.71 &  1.95 & -17.30 \\
     & -0.27 &  1.41 & -1.20 &  1.66 & -2.80 &  1.82 & -16.25 \\
\hline
MgO  & -2.02 &  7.63 & -0.71 & 10.91 & -1.45 & 12.11 &  \\
     & -1.27 &  7.32 & -0.77 & 10.46 & -1.56 & 11.43 &  \\ 
\hline
ZnO-zb & -1.74 &  2.73 & -0.78 &  7.78 & -2.12 &  7.31 & -6.26 \\
       & -0.89 &  2.15 & -0.86 &  7.19 & -2.33 &  6.70 & -6.07 \\
\hline
ZnO-wz & -1.77 &  2.94 &       &       &       &       & -6.03 \\
       & -0.97 &  2.42 &       &       &       &       & -5.94 \\
\hline
ZnS  & -1.17 &  3.38 & -0.85 &  4.73 & -2.16 &  4.64 & -7.56 \\
     & -0.54 &  3.16 & -0.95 &  4.48 & -2.40 &  4.29 & -6.73 \\
\hline
CdS  & -0.99 &  2.09 & -0.78 &  4.14 & -1.91 &  4.54 & -8.66 \\
     & -0.55 &  1.93 & -0.85 &  3.96 & -2.08 &  4.31 & -8.11 \\
\hline\hline
Ar   & -3.60 & 13.20 & -0.17 & 16.36 & -0.50 & 16.06 &  \\
     & -2.68 & 12.54 & -0.19 & 15.75 & -0.56 & 15.41 &  \\
\hline
Ne   & -6.42 & 20.31 & -0.09 & 26.06 & -0.25 & 27.18 &  \\
     & -5.64 & 20.15 & -0.10 & 26.47 & -0.28 & 26.96 &  \\
\hline
LiF  & -3.44 & 14.09 & -0.26 & 15.90 & -1.10 & 20.24 &  \\
     & -2.32 & 13.46 & -0.29 & 15.13 & -1.23 & 20.04 &  \\
\hline\hline
\end{tabular} 
\end{table}
\begin{figure}[tbph]
\begin{center}
\includegraphics[width=0.9\columnwidth]{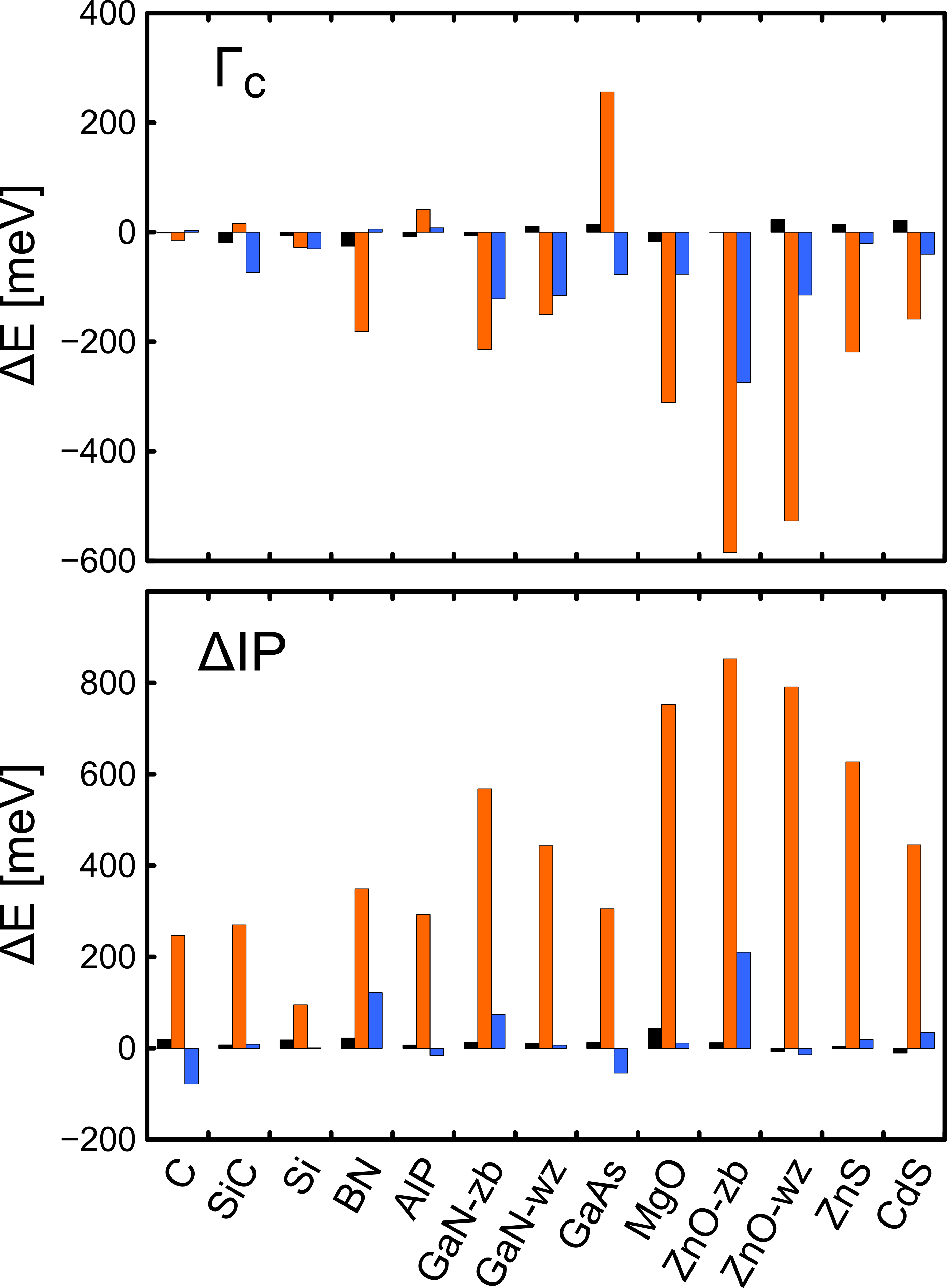}
\end{center}
\caption{\label{fig:gw-deltaE}
Mean absolute deviations ($\Delta E$) of results obtained with the {\it optimized} basis set from those of the {\it default} one (orange bars); results for the direct bandgap at $\Gamma$ (left) and the ionization potential shift (right) based on the extrapolation scheme.
The blue bars indicate differences between this work and Ref.~\onlinecite{Klimes2014a}.
Black bars depict differences between the extrapolation procedure and {\it all-states} calculations.
}
\end{figure}

A comparison of the direct bandgaps (reflected by $\Gamma_c$) for selected compounds and the absolute shifts of the valence band maximum ($\Delta$IP) as calculated with the two basis sets are presented in Fig.~\ref{fig:gw-deltaE} (orange bars).
In the simple mono-atomic semiconductors such as Si and C, the impact of additional local orbitals is rather small (within 30 meV).
This has already been shown earlier in Ref.~\onlinecite{Friedrich2006}.
The basis-set quality turns out significantly more important for all other materials under investigation.
Differences between the two settings are around 200 meV, with the maximum deviation of 0.5 eV, being observerd for ZnO, as expected, in both the zinc-blende and wurtzite phase.
It is also worth noting, that our values obtained using the {\it default} basis configurations are found to be in reasonable agreement with the values published earlier, e.g., in Refs.~\onlinecite{Shishkin2007,Friedrich2010}.
The small influence of the basis set on the QP bandgaps for Si 
and C should not be considered as special.
As already discussed,\cite{Gulans2014} the reason behind is a fortunate error cancellation due to similar convergence behavior of VB and CB.
In general, however, the absolute positions of VB and CB converge rather slowly following the asymptotic law, Eq.~\ref{eq:asymp}.
Therefore, focusing on $\Delta$IP (right panel of Fig.~\ref{fig:gw-deltaE}), rather pronounced differences between the two settings are observed for all compounds.

It is also interesting to compare results from the extrapolation procedure and the {\it all-states} calculations.
The respective differences are presented as black bars in Fig.~\ref{fig:gw-deltaE}.
We find very good overall agreement between the two schemes.
The largest deviation of 25 meV (bandgap of BN) and 43 meV ($\Delta$IP in MgO) could be attributed to the above discussed inherent uncertainty how to choose a proper interval for the extrapolation procedure.

Eventually, we arrive at a comparison of our $G_0W_0$@LDA results to those obtained by Klime\v{s}~{\it et al.}\cite{Klimes2014a} (blue bars in Fig.~\ref{fig:gw-deltaE}).
The agreement between ours and the PAW results is overall good, where the analysis of data presented in Table~\ref{tab:qp-gw} reveals an absolute mean average deviation of 60 meV.
The largest deviations are observed for the transition metal oxides.
The only obvious contradiction is obtained for zinc-blende ZnO, where our bandgap is 270 meV larger than the one of Ref.~\onlinecite{Klimes2014a}.
Good agreement concerning $d$-band positions is found for ZnO, ZnS, and CdS, i.e., in systems where the $d$-states are rather high in energy.
For deep-lying $d$-bands the differences between LAPW+lo and PAW are more pronounced.
Specifically, our values are found to be lower by ~0.24 eV for GaN (-16.11 eV vs -15.87 eV) and ~0.2 eV for GaAs (-17.30 eV vs -17.10 eV).

Unfortunately, in Ref. \onlinecite{Klimes2014a} no data are available for Ar, Ne, and LiF crystals.
With the {\it default} basis sets we reproduce values in good agreement with previous PAW studies.\cite{Shishkin2007} As it follows from Table~\ref{tab:qp-gw}, usage of the {\it optimized} basis sets leads to a drastic increase of the $G_0W_0$ bandgap of LiF, from 13.46 eV to 14.09 eV, that turns out to be much closer to the experimental value of 14.2 eV.

\section{Conclusions}

We have demonstrated a systematic approach how to reach numerically accurate $G_0W_0$ quasiparticle energies within the full-potential linearized-augmented planewaves method.
The decisive role of local orbitals has been reexamined.
Comparing different basis sets ({\it optimized} vs {\it default}), we conclude that the basis-set incompleteness error is a main reason for the dispersion among $G_0W_0$ results observed in literature.
We have analyzed in detail how the completeness of the LAPW+lo basis impacts the $G_0W_0$ results.
On the one hand, the addition of local orbitals leads to the asymptotically correct behavior of the quasiparticle band energies in the limit of large number of empty states.
This behavior, which is common for all plane-wave codes, suggests a simple and efficient extrapolation procedure to evaluate the band energies corresponding to the summation over infinite unoccupied states.
On the other hand, and in contrast to any method based on pseudization, the easily adjustable local-orbital 
part of the LAPW+lo basis provides an efficient control over the basis-set quality which is important for excited-state calculations.
It turns out that local orbitals provide additional degrees of freedom that are essential for resolving the short-range features of the linear response in the regions where it matters most.
These regions are enclosed by muffin tins and correspond to the part of the unit cell with the highest density.
Monitoring the convergence of $G_0W_0$ results with respect to the LAPW+lo basis size, we have elaborated an alternative to the extrapolation procedure, with the results obtained from a single calculation where all available states are included ({\it all-states} summation).
Applying both approaches, the $G_0W_0$ quasiparticle band structures have been computed for a test set of small- and wide-gap semiconductors, as well as insulators.
We found very good agreement between values obtained by both computational approaches which insures a high quality of our results.
Comparing our 
results with that recently obtained by Kresse and coworkers~\cite{Klimes2014a}, we confirm the overall good agreement between FLAPW and PAW $G_0W_0$ quasiparticle energies for a larger set of materials.

\begin{acknowledgments}
Financial support from the German Science Foundation (DFG), SFB 951 and SFB 568, is appreciated.
\end{acknowledgments}

\appendix
\section{\label{app:losetups}Local-orbital configurations}

\begin{table}[htb]
\caption{\label{tab:lo} Radial part of local orbitals for Zn and O atoms designed for groundstate calculations.
The prefactors $a$ and $b$ have symbolic meaning and are different for each local orbital.
These lo's represent the {\it default} setting.}
\begin{tabular}{l|l|l}
  \hline\hline
  $l$ & Zn & O \\
  \hline
$s$ &     $a \; \! u_0(r;\epsilon_{4s})+ b \; \! \dot{u}_0(r;\epsilon_{4s})$  &   $a \; \! u_0(r;\epsilon_{2s})+ b \; \!\dot{u}_0(r;\epsilon_{2s})$ \\
    &     $a \; \! u_0(r;\epsilon_{4s})+ b \; \! \ddot{u}_0(r;\epsilon_{4s})$ &   $a \; \! u_0(r;\epsilon_{2s})+ b \; \!\ddot{u}_0(r;\epsilon_{2s})$ \\
    &     $a \; \! u_0(r;\epsilon_{4s})+ b \; \! u_0(r;\epsilon_{3s})$        &   \\
    &     $a \; \! u_0(r;\epsilon_{3s})+ b \; \! \dot{u}_0(r;\epsilon_{3s})$  &   \\
\hline
$p$ &     $a \; \! u_1(r;\epsilon_{4p})+ b \; \! \dot{u}_1(r;\epsilon_{4p})$  &   $a \; \! u_1(r;\epsilon_{2p})+ b \; \!\dot{u}_1(r;\epsilon_{2p})$ \\
    &     $a \; \! u_1(r;\epsilon_{4p})+ b \; \! \ddot{u}_1(r;\epsilon_{4p})$ &   $a \; \! u_1(r;\epsilon_{2p})+ b \; \! \ddot{u}_1(r;\epsilon_{2p})$ \\
    &     $a \; \! u_1(r;\epsilon_{4p})+ b \; \! u_1(r;\epsilon_{3p})$        &   \\
    &     $a \; \! u_1(r;\epsilon_{3p})+ b \; \!\dot{u}_1(r;\epsilon_{3p})$  &   \\

          \hline
$d$ &    $a \; \! u_2(r;\epsilon_{3d})+ b \; \!\dot{u}_2(r;\epsilon_{3d})$  &    \\
    &    $a \; \! u_2(r;\epsilon_{3d})+ b \; \!\ddot{u}_2(r;\epsilon_{3d})$ &    \\
\hline\hline
\end{tabular} 
\end{table}

\begin{table}[htbp]
\caption{\label{tab:xslo} Radial part of local orbitals for Zn and O atoms designed for excited-state calculations.
The prefactors $a$ and $b$ have symbolic meaning and are different for each local orbital.
Together with the lo's of Table \ref{tab:lo}, they represent the {\it optimized} setting.}
\begin{tabular}{l|l|l}
  \hline\hline
  $l$ & Zn & O \\
  \hline
$s$ &     $a \; \! u_0(r;\epsilon_{4s})+ b \; \! u_0(r;\epsilon_{5s})$  &   $a \; \! u_0(r;\epsilon_{2s})+ b \; \! u_0(r;\epsilon_{3s})$ \\
    &     $a \; \! u_0(r;\epsilon_{4s})+ b \; \! u_0(r;\epsilon_{6s})$  &   $a \; \! u_0(r;\epsilon_{2s})+ b \; \! u_0(r;\epsilon_{4s})$ \\
    &     $a \; \! u_0(r;\epsilon_{4s})+ b \; \! u_0(r;\epsilon_{7s})$  &   $a \; \! u_0(r;\epsilon_{2s})+ b \; \! u_0(r;\epsilon_{5s})$ \\
    &                                                       &   $a \; \! u_0(r;\epsilon_{2s})+ b \; \! u_0(r;\epsilon_{6s})$ \\
\hline
$p$ &     $a \; \! u_1(r;\epsilon_{4p})+ b \; \! u_1(r;\epsilon_{5p})$  &   $a \; \! u_1(r;\epsilon_{2p})+ b \; \! u_1(r;\epsilon_{3p})$ \\
    &     $a \; \! u_1(r;\epsilon_{4p})+ b \; \! u_1(r;\epsilon_{6p})$  &   $a \; \! u_1(r;\epsilon_{2p})+ b \; \! u_1(r;\epsilon_{4p})$ \\
    &     $a \; \! u_1(r;\epsilon_{4p})+ b \; \! u_1(r;\epsilon_{7p})$  &   $a \; \! u_1(r;\epsilon_{2p})+ b \; \! u_1(r;\epsilon_{5p})$ \\
    &     $a \; \! u_1(r;\epsilon_{4p})+ b \; \! u_1(r;\epsilon_{8p})$  &   $a \; \! u_1(r;\epsilon_{2p})+ b \; \! u_1(r;\epsilon_{6p})$ \\
    &                                                       &   $a \; \! u_1(r;\epsilon_{2p})+ b \; \! u_1(r;\epsilon_{7p})$ \\
          \hline
$d$ &     $a \; \! u_2(r;\epsilon_{3d})+ b \; \! u_2(r;\epsilon_{4d})$  &   $a \; \! u_2(r;\epsilon_{3d})+ b \; \!\dot{u}_2(r;\epsilon_{3d})$ \\
    &     $a \; \! u_2(r;\epsilon_{3d})+ b \; \! u_2(r;\epsilon_{5d})$  &   $a \; \! u_2(r;\epsilon_{3d})+ b \; \! u_2(r;\epsilon_{4d})$ \\
    &     $a \; \! u_2(r;\epsilon_{3d})+ b \; \! u_2(r;\epsilon_{6d})$  &   $a \; \! u_2(r;\epsilon_{3d})+ b \; \! u_2(r;\epsilon_{5d})$ \\
    &     $a \; \! u_2(r;\epsilon_{3d})+ b \; \! u_2(r;\epsilon_{7d})$  &   $a \; \! u_2(r;\epsilon_{3d})+ b \; \! u_2(r;\epsilon_{6d})$ \\
    &     $a \; \! u_2(r;\epsilon_{3d})+ b \; \! u_2(r;\epsilon_{8d})$  &   $a \; \! u_2(r;\epsilon_{3d})+ b \; \! u_2(r;\epsilon_{7d})$ \\
    &                                                       &   $a \; \! u_2(r;\epsilon_{3d})+ b \; \! u_2(r;\epsilon_{8d})$ \\    
    &                                                       &   $a \; \! u_2(r;\epsilon_{3d})+ b \; \! u_2(r;\epsilon_{9d})$ \\        
\hline    
$f$ &     $a \; \! u_3(r;\epsilon_{4f})+ b \dot{u}_3(r;\epsilon_{4f})$  &   $a \; \! u_3(r;\epsilon_{4f})+ b \; \! \dot{u}_3(r;\epsilon_{4f})$ \\
    &     $a \; \! u_3(r;\epsilon_{4f})+ b \; \! u_3(r;\epsilon_{5f})$  &   $a \; \! u_3(r;\epsilon_{4f})+ b \; \! u_3(r;\epsilon_{5f})$ \\
    &     $a \; \! u_3(r;\epsilon_{4f})+ b \; \! u_3(r;\epsilon_{6f})$  &   $a \; \! u_3(r;\epsilon_{4f})+ b \; \! u_3(r;\epsilon_{6f})$ \\
    &     $a \; \! u_3(r;\epsilon_{4f})+ b \; \! u_3(r;\epsilon_{7f})$  &   $a \; \! u_3(r;\epsilon_{4f})+ b \; \! u_3(r;\epsilon_{7f})$ \\
    &     $a \; \! u_3(r;\epsilon_{4f})+ b \; \! u_3(r;\epsilon_{8f})$  &   $a \; \! u_3(r;\epsilon_{4f})+ b \; \! u_3(r;\epsilon_{8f})$ \\
    &     $a \; \! u_3(r;\epsilon_{4f})+ b \; \! u_3(r;\epsilon_{9f})$  &   $a \; \! u_3(r;\epsilon_{4f})+ b \; \! u_3(r;\epsilon_{9f})$ \\    
    &     $a \; \! u_3(r;\epsilon_{4f})+ b \; \! u_3(r;\epsilon_{10f})$  &  $a \; \! u_3(r;\epsilon_{4f})+ b \; \! u_3(r;\epsilon_{10f})$  \\    
\hline\hline
\end{tabular} 
\end{table}

Below we provide all important details on our way of generating an {\it optimized} basis for describing unoccupied states within the LAPW+lo formalism.
As an illustration, we consider the setup for wurtzite ZnO.
As a first step, we produce a set of local orbitals suited for ground-state calculations.
These local orbitals, as listed in Tab.~\ref{tab:lo}, employ $u_{l\alpha}$, $\dot{u}_{l\alpha}$ and $\ddot{u}_{l\alpha}$ as $f_\mu$ and $g_\mu$ used in Eq.~\ref{eq:lo}.

For Zn, the radial functions of local orbitals with $l=0$ contain functions obtained for the energy parameters $\epsilon_{3s}$ and $\epsilon_{4s}$, as we consider $3s$, and likewise $3p$, as a semi-core states.
All energy parameters used for the local orbitals in Tab.~\ref{tab:lo} are close to the corresponding KS eigenvalues.

$G_0W_0$ calculations involve a range of bands that enter Eqs.~\ref{eq:gw:G0} and \ref{eq:gw:p0}, but the basis set discussed above describes well only semicore, valence, and a few low-lying conduction bands.
The additional local orbitals are constructed in the form
\begin{equation}
 \label{eq:xslo}
 \phi^\mathrm{lo}(\mathbf{r})=\left[a \; u_l(r;\epsilon_{l}^v)+ b \; u_l(r;\epsilon_{l}^{c})\right] \;Y_{lm}(\hat{\mathbf{r}}),
\end{equation}
where the energy parameter $\epsilon_{l}^v$ corresponds to the respective valence bands with the particular $l$-character.
The energy parameters, $\epsilon_{l}^{c}$, representing unoccupied states of one $l$ channel must be chosen such that the corresponding radial functions $u_l(r;\epsilon_{l}^{c})$ exhibit a different number of nodes in the $MT$ region.
To do so, we pick $\epsilon_{l}^{c}$ according to the Wigner-Seitz rules~\cite{Andersen1973}, according to which the bottom $\epsilon_b$ and top $\epsilon_t$ of a band with a predominantly $l$-character satisfy the conditions at the muffin-tin border $R_{\mathrm{MT}}$.
\begin{align}
    \mathrm{d}u_{l\alpha}(R_\mathrm{MT},\epsilon_\mathrm{b})/\mathrm{d}E &= 0 \\
	   u_{l\alpha} \left( R_\mathrm{MT},\epsilon_\mathrm{t} \right) &= 0.
\end{align}
Setting $\epsilon_{l}^{c}=(\epsilon_b+\epsilon_t)/2$, allows us to generate local orbitals with a given number of nodes.
 
For sufficiently large $l$, there is no valence state with this particular $l$-character.
In this case, we still use Eq.~\ref{eq:xslo}, albeit with $\epsilon_{l}^v$ replaced by an approximate Fermi energy $\epsilon_\mathrm{F}$, and set up an
additional local orbital as
\begin{equation}
 \label{eq:hil}
 \phi^\mathrm{lo}(\mathbf{r})=\left[ a \; u_l(r;\epsilon_\mathrm{F})+ b \; \dot{u}_l(r;\epsilon_\mathrm{F}) \right] \; Y_{lm}(\hat{\mathbf{r}}).
\end{equation}

\begin{table}[htbp]
\caption{\label{tab:energies} Energy parameters used in Tabs.~\ref{tab:lo} and \ref{tab:xslo}.
The zero energy corresponds to the valence-band maximum.
All values are given in Hartrees.}
\begin{tabular}{c|rr|| c|rr|| c|rr}
  \hline\hline
   & Zn & O & & Zn & O & & Zn & O \\
   \hline
 $\epsilon_{2s}$ &   --~~ & $-0.7$ & $\epsilon_{2p}$ &   --~~ &  0.0 & $\epsilon_{3d}$ & $-0.3$ &  0.0 \\
 $\epsilon_{3s}$ & $-4.5$ &  16.4  & $\epsilon_{3p}$ & $-2.9$ & 14.2 & $\epsilon_{4d}$ &   14.0 &  3.2 \\
 $\epsilon_{4s}$ &   0.0  &  31.6  & $\epsilon_{4p}$ &   0.0  & 27.6 & $\epsilon_{5d}$ &   27.8 & 10.5 \\
 $\epsilon_{5s}$ &  29.1  &  50.9  & $\epsilon_{5p}$ &  13.8  & 45.1 & $\epsilon_{6d}$ &   45.8 & 21.9 \\
 $\epsilon_{6s}$ &  49.6  &  74.2  & $\epsilon_{6p}$ &  29.1  & 66.5 & $\epsilon_{7d}$ &   67.9 & 37.4 \\
 $\epsilon_{7s}$ &  74.5  &   --~~ & $\epsilon_{7p}$ &  48.8  & 91.9 & $\epsilon_{8d}$ &   94.0 & 56.9 \\ 
                 &        &        & $\epsilon_{8p}$ &  72.7  & --~~ & $\epsilon_{9d}$ &   --~~ & 80.3 \\
\hline                 
 $\epsilon_{4f}$ &    0.0 &  0.0   & $\epsilon_{5g}$ &    0.0 &  0.0 & $\epsilon_{6h}$ &    0.0 &  0.0 \\
 $\epsilon_{5f}$ &    4.6 &  6.0   & $\epsilon_{6g}$ &    8.0 &  9.1 & $\epsilon_{7h}$ &   11.7 & 12.6 \\
 $\epsilon_{6f}$ &   12.4 & 15.7   & $\epsilon_{7g}$ &   18.8 & 12.6 & $\epsilon_{8h}$ &   25.1 & 26.9 \\
 $\epsilon_{7f}$ &   24.3 & 29.3   & $\epsilon_{8g}$ &   33.2 & 36.8 & $\epsilon_{9h}$ &   41.9 & 44.8 \\
 $\epsilon_{8f}$ &   40.3 & 46.9   & $\epsilon_{9g}$ &   51.6 & 56.5 & $\epsilon_{10h}$&   62.5 & 66.4 \\
 $\epsilon_{9f}$ &   60.5 & 68.4   & $\epsilon_{10g}$&   74.0 & 80.0 & $\epsilon_{11h}$&   87.1 & 92.0 \\ 
 $\epsilon_{10f}$&   84.7 & 93.8   & $\epsilon_{11g}$&   --~~ & --~~ & $\epsilon_{12h}$&   --~~ & --~~ \\                 
\hline                 
 $\epsilon_{7i}$  &    0.0 &  0.0   & $\epsilon_{8j}$ &    0.0 &  0.0 & $\epsilon_{9k}$ &    0.0 &  0.0 \\
 $\epsilon_{8i}$  &   15.7 & 16.5   & $\epsilon_{9j}$ &   20.1 & 20.8 & $\epsilon_{10k}$ &   24.9 & 25.6 \\
 $\epsilon_{10i}$ &   31.6 & 33.2   & $\epsilon_{10j}$ &   38.4 & 39.8 & $\epsilon_{11k}$ &   45.7 & 47.0 \\
 $\epsilon_{11i}$ &   50.7 & 53.1   & $\epsilon_{11j}$ &   59.8 & 61.9 & $\epsilon_{12k}$ &   69.3 & 71.1 \\
 $\epsilon_{12i}$ &   73.6 & 76.8   & $\epsilon_{12j}$ &   84.8 & 87.6 & $\epsilon_{13k}$&   96.5 & 98.9 \\
 \hline\hline
\end{tabular} 
\end{table}

\section{\label{app:crystals}Crystal structures}

The crystal structures and lattice parameters, as well as the actual muffin-tin radii, $R_{\mathrm{MT}}$, for all systems calculated in this work are specified in Table~\ref{tab:crystals}. For all binary compounds, we use equal $MT$ radii for both species. The deep lying $3s$ and $3p$ states of Zn, Ga, and As atoms are considered as valence states in order to minimize core-charge leakage.

\begin{table}[tbhp]
  \caption{Crystal structures and experimental lattice parameters as used in the present work (adopted from Refs.~\onlinecite{Gruneis2014,Shishkin2007}).
In addition, the $MT$ radii are specified.
In case of binary materials, we use equal $R_{\mathrm{MT}}$ for both species.}
  \label{tab:crystals}
  \begin{tabular}{l l c c c c}
	\hline \hline
  Material  & structure type & a [\AA] & c [\AA] & u & $R_{\mathrm{MT}}$ \\
  \hline        
  C     & diamond    & 3.567     & -         & -      & 1.4 \\
  SiC   & zincblende & 4.358     & -         & -      & 1.6 \\
  Si    & diamond    & 5.431     & -         & -      & 2.1 \\
  BN    & zincblende & 3.616     & -         & -      & 1.4 \\
  AlP   & zincblende & 5.463     & -         & -      & 2.2 \\
  GaN   & zincblende & 4.535     & -         & -      & 1.6 \\
        & wurtzite   & 3.190     & 5.186     & 0.3789 & 1.6 \\
  GaAs  & zincblende & 5.654     & -         & -      & 2.2 \\
  MgO   & rocksalt   & 4.211     & -         & -      & 1.6 \\
  ZnO   & zincblende & 4.584     & -         & -      & 1.6 \\
        & wurtzite   & 3.250     & 5.207     & 0.3819 & 1.6 \\
  ZnS   & zincblende & 5.409     & -         & -      & 2.2 \\
  CdS   & zincblende & 5.818     & -         & -      & 2.2 \\
  Ar    & fcc        & 5.260     & -         & -      & 3.0 \\
  Ne    & fcc        & 4.430     & -         & -      & 2.8 \\
  LiF   & rocksalt   & 4.010     & -         & -      & 1.8 \\
	\hline \hline
  \end{tabular}
\end{table}

\section{\label{app:kqtest} {Convergence of k and q-point grids}}

\begin{figure}[htbp]
  \centering
  \includegraphics[width=0.7\columnwidth]{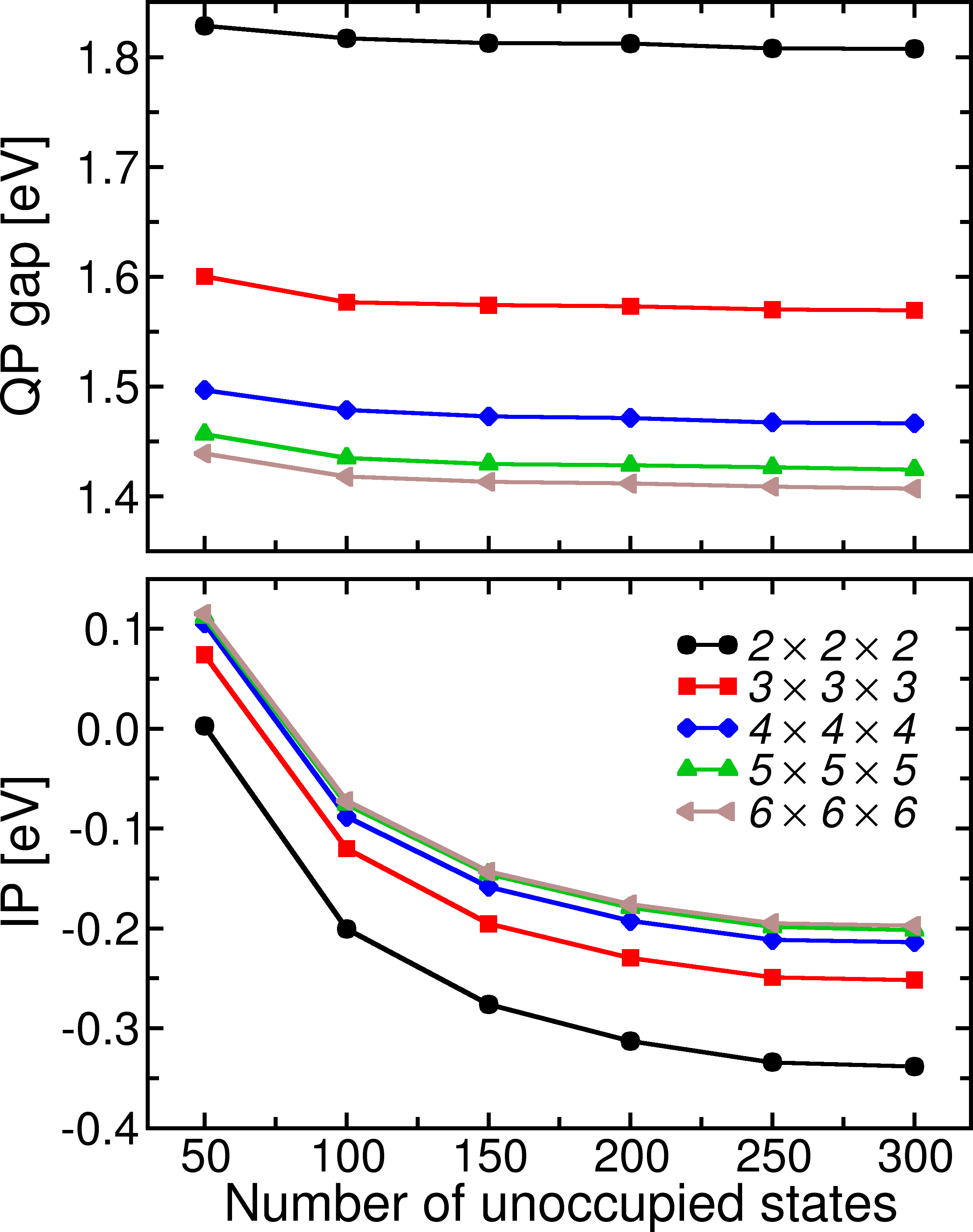}
  \caption{\label{fig:kqtest}Convergence of the QP gap (top) and absolute shift of the VB maximum compared to LDA calculations (bottom) in GaAs as a function of the number of empty states for different  $\mathbf{k}/\mathbf{q}$-point grids.}
\end{figure}

Fig.~\ref{fig:kqtest} presents convergence studies for the quasiparticle bandgap (top) and the position of VB (bottom) in GaAs concerning the $\mathbf{k}/\mathbf{q}$-point grid.
Changing the grid size results in an almost rigid shift of both curves, $E_g(N)$ and IP($N$).
This observation suggests that one can perform convergence tests with respect to $\mathbf{k}/\mathbf{q}$-point size independently from the corresponding test with respect to the number of unoccupied states.



\end{document}